\documentclass[12pt]{article}
\usepackage{amsmath,amssymb,epsfig}
\makeatletter \@addtoreset{equation}{section} \makeatother
\renewcommand{\theequation}{\thesection.\arabic{equation}}
\newcommand{\ba}{\begin{array}}
\newcommand{\ea}{\end{array}}
\newcommand{\beq}{\begin{equation}}
\newcommand{\eeq}{\end{equation}}
\newcommand{\bea}{\begin{eqnarray}}
\newcommand{\eea}{\end{eqnarray}}

\def\bce{\begin{center}}
\def\ece{\end{center}}

\def\nonu{\nonumber}

\def\pa{\partial}
\def\al{\alpha}
\def\be{\beta}

\def\de{\delta}

\def\la{\lambda}

\def\diag{\mathop{\rm diag}}

\def\eps6{{\displaystyle \mathop{\epsilon}^{6}}{}}

\def\nab6{{\displaystyle \mathop{\nabla}^{6}}{}}

\def\0{{\sst{(0)}}}
\def\1{{\sst{(1)}}}
\def\2{{\sst{(2)}}}
\def\3{{\sst{(3)}}}
\def\4{{\sst{(4)}}}
\def\5{{\sst{(5)}}}
\def\6{{\sst{(6)}}}
\def\7{{\sst{(7)}}}
\def\8{{\sst{(8)}}}

\def\ba{\begin{array}}
\def\ea{\end{array}}
\def\beq{\begin{equation}}
\def\eeq{\end{equation}}
\def\be{\begin{equation}}
\def\ee{\end{equation}}

\def\Tr{\mathop{\rm Tr}}

\def\diag{\mathop{\rm diag}}

\def\la{\lambda}
\def\eps{\epsilon}

\def\ba{\begin{array}}
\def\ea{\end{array}}
\def\beq{\begin{equation}}
\def\eeq{\end{equation}}
\def\be{\begin{equation}}
\def\ee{\end{equation}}

\def\Tr{\mathop{\rm Tr}}

\def\diag{\mathop{\rm diag}}

\def\la{\lambda}
\def\eps{\epsilon}

\newcommand{\bean}{\begin{eqnarray*}}
\newcommand{\eean}{\end{eqnarray*}}

\begin{document}
\thispagestyle{empty} \addtocounter{page}{-1}
   \begin{flushright}
\end{flushright}

\vspace*{1.3cm}

 \centerline{ \Large \bf  Towards Holographic Gravity Dual of } 
\vspace{.3cm} 
\centerline{ \Large \bf  
${\cal N}=1$ Superconformal Chern-Simons Gauge Theory   } 
\vspace*{1.5cm}
\centerline{{\bf Changhyun Ahn }
} 
\vspace*{1.0cm} 
\centerline{\it  
Department of Physics, Kyungpook National University, Taegu
702-701, Korea} 
\vspace*{0.8cm} 
\centerline{\tt ahn@knu.ac.kr
} 
\vskip2cm

\centerline{\bf Abstract}
\vspace*{0.5cm}

As we re-examine the known holographic ${\cal N}=1$
supersymmetric renormalization group flow in four dimensions, 
we describe the mass-deformed Bagger-Lambert theory or 
equivalently the mass-deformed $U(2) \times
U(2)$ Chern-Simons gauge theory with level $k=1$ or $2$, 
that has $G_2$
symmetry, by adding a mass term
for one of the eight adjoint superfields. 
We obtain a detailed correspondence between the fields of
$AdS_4$ supergravity and composite operators of the 
infrared field theory in three dimensions.
The geometric superpotential from an eleven dimensional viewpoint
is obtained for M2-brane probe analysis.

\baselineskip=18pt
\newpage
\renewcommand{\theequation}
{\arabic{section}\mbox{.}\arabic{equation}}

\section{Introduction}

When one reduces 11-dimensional
supergravity theory to four-dimensional ${\cal N}=8$ gauged supergravity, 
the four-dimensional
spacetime is warped by warp factor that depends on both four-dimensional
coordinates and 7-dimensional internal coordinates.   
This warp factor provides an understanding of the different relative scales
of the 11-dimensional solutions corresponding to 
the critical points in ${\cal N}=8$ gauged supergravity. 
An important aspect of the holographic
duals \cite{Maldacena} is the notion that the
radial coordinate of $AdS_4$ can be viewed as a measure of energy.
A supergravity kink description interpolating between 
$r \rightarrow \infty$ 
and $r \rightarrow -\infty$ can be interpreted as an explicit construction 
of the renormalization group(RG) flow
between the ultraviolet(UV) fixed point and 
the infrared(IR) fixed point of the three dimensional boundary 
field theory. 

It is known \cite{Warner83} that there exist five nontrivial critical points
for the scalar potential of gauged ${\cal N}=8$ supergravity:
$SO(7)^{+}, SO(7)^{-}, G_2, SU(4)^{-}$ and $SU(3) \times U(1)$. 
Among them $G_2$-invariant 7-ellipsoid and $SU(3) \times U(1)$-invariant
stretched 7-ellipsoid are stable and supersymmetric.
The holographic
RG flow equations from ${\cal N}=8$ $SO(8)$-invariant UV fixed point 
to ${\cal N}=2$ $SU(3) \times U(1)$-invariant IR fixed point
were constructed in \cite{AP}.
Moreover,  
the holographic
RG flow equations from ${\cal N}=8$ $SO(8)$-invariant UV fixed point 
to
${\cal N}=1$ $G_2$-invariant IR 
fixed point were obtained in \cite{AW,AI02}. See also \cite{AR99}. 
An exact solution to the 11-dimensional 
bosonic equations corresponding to the $M$-theory lift of the ${\cal N}=2$ 
$SU(3) \times U(1)$-invariant RG flow was found in \cite{CPW} and 
its Kahler structure was extensively studied in \cite{JLP01}. 
Furthermore, the $M$-theory lift of the ${\cal N}=1$ $G_2$-invariant 
RG flow was done in \cite{AI02}.

Bagger and Lambert(BL) proposed a Lagrangian to describe 
the low energy dynamics of multiple M2-branes in \cite{BL0711}.
See also related papers \cite{Gustavsson07}.
This BL theory 
is three dimensional ${\cal N}=8$ supersymmetric theory with $SO(8)$ 
global symmetry 
based on new 3-algebra
and this 3-algebra with Lorentzian signature was proposed
by \cite{GMR}.
The generators of the 3-algebra consist of the generators of an arbitrary
semisimple Lie algebra and two additional null generators.

Very recently, in \cite{ABJM}, 
in order to have arbitrary number of M2-branes,
three dimensional Chern-Simons matter theories
with gauge group $U(N) \times U(N)$ and level $k$ 
which have ${\cal N}=6$
superconformal symmetry are constructed.
They describe this theory as the low energy limit of $N$ M2-branes at 
${\bf C}^4/{\bf Z}_k$ singularity.
In particular, when $N=2$, this leads to the BL theory.
Furthermore, the full $SU(4)_R$ symmetry of \cite{ABJM} is proved 
explicitly in \cite{BKKS}.
By examining the holographic ${\cal N}=2$
supersymmetric renormalization group flow solution among  
five nontrivial critical points above, in four dimensions, 
the mass-deformed BL theory that has $SU(3)_I \times U(1)_R$
symmetry is studied in \cite{Ahn08} by the addition of mass term
for one of the four adjoint chiral superfields. 
We list some relevant works on 
the M2-brane theory in \cite{GGY}-\cite{MuPa}. 

Then it is natural to ask what happens 
for the holographic description with 
stable and ${\cal N}=1$ supersymmetric 
$G_2$-invariant 7-ellipsoid compactification by deforming 
three dimensional Chern-Simons matter theories
with gauge group $U(N) \times U(N)$ and level $k$?
As a first step, we consider 
the $U(2) \times U(2)$
Chern-Simons gauge theory of
\cite{ABJM} with level $k=1$ or $k=2$ which preserves $G_2$ 
global symmetry.

In this paper, 
starting from the first order differential equations, that are the
supersymmetric flow solutions in four dimensional ${\cal N}=8$ gauged
supergravity
interpolating between an exterior $AdS_4$ region with maximal
${\cal N}=8$ supersymmetry
and an interior $AdS_4$ with one eighth(i.e., ${\cal N}=1$) 
of the maximal supersymmetry,  
we would like to interpret this as the RG flow in BL theory
\footnote{When we describe BL theory in this paper, the two 
M2-branes theory($N=2$) is equivalent to $U(2) \times U(2)$
Chern-Simons gauge theory of
\cite{ABJM} with level $k=1$ or $k=2$.  }
which has $OSp(8|4)$ symmetry
broken to the deformed BL theory which has $OSp(1|4)$ symmetry
by the addition of a mass term for
one of the eight adjoint superfields. 

An exact correspondence 
is obtained between fields of bulk supergravity in the $AdS_4$ region
in four dimensions 
and composite operators of the IR field theory in three dimensions. 
It is easy to check how the supersymmetry
breaks for specific deformation and one can extract the correct full
superpotential including the superpotential before the deformation 
also. 
The three dimensional analog of Leigh-Strassler 
RG flow in mass-deformed BL theory in three dimensions is expected 
by looking at its holographic dual theory in four dimensions along the
line of \cite{Ahn08}.   

In section 2, we review the supergravity solution in four dimensions
in the context of RG flow, describe two supergravity critical points 
and present the supergravity multiplet in terms of 
$G_2$ invariant ones. The decomposition of $SO(8)$ into $G_2$ is 
presented.
 
In section 3, we deform BL theory by adding one of the mass term among
eight superfields, along the lines of \cite{HLL}, write down
the $SO(7)^{+}$-invariant 
superpotential in ${\cal N}=1$ superfields which will be invariant
under the $G_2$ after integrating out the massive superfield 
and describe the scale
dimensions for the superfields at UV.  

In section 4, 
the $OSp(1|4)$ representations(energy and spin) and 
$G_2$ representations
in the supergravity mass spectrum for
each multiplet at
the ${\cal N}=1$ critical point and the corresponding ${\cal N}=1$
superfield in the boundary gauge theory are given.
For this computation, the 11-dimensional ``geometric'' 
superpotential which reduces to the usual superpotential for the
particular internal coordinate is needed to analyze the M2-brane 
analysis.

In section 5, 
we end up with the future directions.

\section{The holographic ${\cal N}=1$ 
supersymmetric RG flow in four dimensions }

In ${\cal N}=8$ supergravity \cite{dN82}, there exists 
an ${\cal N}=1$ supersymmetric $G_2$-invariant vacuum \cite{dNW}. 
To arrive at this critical point, one should turn on expectation 
values of both 
scalar and pseudo-scalar fields 
where the 
completely antisymmetric self-dual and anti-self-dual
tensors are invariant under $G_2$ since the $G_2$ is the common
subgroup of $SO(7)^{+}$ acting on the scalar and $SO(7)^{-}$ acting on
the pseudo-scalar respectively.
Then the 56-bein can be written as $ 56\times 56$ matrix whose
elements are some functions of these scalar and pseudo-scalars. 
Then the $G_2$-invariant 
scalar potential of ${\cal N}=8$ supergravity in terms of the original
variables of \cite{dNW} is given by, through the superpotential found
in \cite{AW,AI02}, 
\bea
V(\la, \al) & = & g^2 \left[ \frac{16}{7} \left(
\frac{\pa W}{\pa \la} \right)^2 + \frac{2}{7p^2 q^2}
\left(
\frac{\pa W}{\pa \al} \right)^2  - 6 \,W^2 \right],
\label{pot}
\eea
where $g$ is a coupling of the theory, 
we introduce the hyperbolic functions of $\lambda$ as 
\bea
p = \cosh \left( \frac{\la}{2\sqrt{2}} \right), \qquad
q = \sinh \left( \frac{\la}{2\sqrt{2}} \right)
\label{pq}
\eea
and the superpotential, which can be read off from the element of $A_1$
tensor of the theory, is a magnitude of complex function on the
variables $\la$ and $\al$ 
\bea
W(\la, \al) & = & | p^7+e^{7i\al} q^7 +7 \left( p^3 q^4 e^{4i\al} +
p^4 q^3 e^{3i\al}\right) |.
\label{W}
\eea

There exist two critical points and let us summarize these in Table 1.

$\bullet$ $SO(8)$ critical point

This is well-known, trivial critical point at which the $\la$ field
vanishes with arbitrary $\al$ 
and whose cosmological constant $\Lambda=-6g^2$ 
from (\ref{pot}) and which 
preserves ${\cal N}=8$ supersymmetry. 

$\bullet$ $G_2$ critical point

There is a critical point at
$\la=\sqrt{2}\,\sinh^{-1}\!\sqrt{\frac{2}{5}(\sqrt{3}-1)}$
and $\al=\cos^{-1}\frac{1}{2}\sqrt{3-\sqrt{3}}$
and the cosmological constant $\Lambda=-\frac{216 
\sqrt{2}}{25 \sqrt{5}}\,3^{1/4} g^2
$.
This critical point has an unbroken ${\cal N}=1$ supersymmetry.

\begin{table} 
\begin{center}
\begin{tabular}{|c|c|c|c|c|} \hline
Symmetry
& $\la$  & $\al$ & V & W  \\
\hline
   $SO(8)$ & $0$ & \mbox{any}  & $- 6\,g^2$ & $1$ \\
 $G_2$ & $\sqrt{2}\,\sinh^{-1}\!\sqrt{\frac{2}{5}(\sqrt{3}-1)}$ &
 $\cos^{-1}\frac{1}{2}\sqrt{3-\sqrt{3}}$
 & $-\frac{216 \sqrt{2}}{25 \sqrt{5}}\,3^{1/4} g^2$ 
 & $\sqrt{\frac{36 \sqrt{2}\,3^{1/4}}{25 \sqrt{5}}}$  \\
\hline
\end{tabular} 
\end{center}
\caption{\sl 
Summary of two critical points with symmetry group, supergravity
fields($\la$ and $\al$), scalar potential($V$) and superpotential($W$).}
\end{table} 

For the supergravity description of the 
nonconformal RG flow from one scale to 
another connecting these two critical points, 
the three dimensional Poincare invariant metric has the form 
$
ds^2= e^{2A(r)} \eta_{\mu' \nu'} dx^{\mu'} dx^{\nu'} + dr^2$
where $\eta_{\mu' \nu'}=(-,+,+)$
and $r$ is the coordinate transverse to the domain wall.
Then the supersymmetric flow equations \cite{AW,AI02} with 
(\ref{W}) and (\ref{pq})
are 
described as
\bea 
\frac{d \la}{d r} & = &
\frac{8\sqrt{2}}{7}\,g \,\partial_{\la} W ,\qquad
\frac{d \alpha}{d r}  =  
\frac{\sqrt{2}}{7p^2 q^2}\,g \,\pa_{\al} W, \qquad 
\frac{d A}{d r}  =  -\sqrt{2} \,g \,W.
\label{flow}
\eea
We'll come back these flow equations when we discuss about the scalar function
at IR in section 4.

The fields of the ${\cal N}=8$ theory
transforming in $SO(8)$ representations should be 
decomposed into $G_2$ representations. 
According to the decomposition $SO(8) 
\rightarrow G_2$ given in (\ref{so8g2}) 
which can be obtained from the branching rules 
of $SO(8) \rightarrow SO(7)$ by (\ref{so8so7}) 
and $SO(7) \rightarrow G_2$ by (\ref{so7g2}) we present in
the appendix A, 
the spin $\frac{3}{2}$ field breaks 
into a singlet and a fundamental of $G_2$
\bea
\mbox{Spin} \; \frac{3}{2}: 
\qquad {\bf 8} \rightarrow [{\bf 1}] \oplus {\bf  7}
\label{8}
\eea
under  $SO(8) 
\rightarrow G_2$
and
the singlet in square bracket
corresponds to the massless graviton of the ${\cal N}=1$ theory.
From the further branching rule of $G_2 \rightarrow SU(3)$ by (\ref{g2su3})
one sees that 
the above septet$({\bf 7})$ which will be located at ${\cal N}=1$ 
massive gravitino multiplet breaks into triplet$({\bf 3})$ and
anti-triplet$({\bf \bar{3}})$ that enter into the component of 
${\cal N}=2$ short massive gravitino multiplet
as well as a singlet$({\bf 1})$ which enters into 
the component of ${\cal N}=2$ massless graviton multiplet 
under the $G_2 \rightarrow SU(3)$ breaking.
Although the global symmetry $G_2$ is reduced to $SU(3)$, the ${\cal
  N}=1$ 
supersymmetry
is enhanced to ${\cal N}=2$ supersymmetry.

According to the decomposition $SO(8) 
\rightarrow G_2$ given in (\ref{so8g2}), one obtains
the following decomposition by a singlet, two fundamentals, adjoint
and a symmetric representation
\bea
\mbox{Spin} \; \frac{1}{2}: \qquad
{\bf 56} & \rightarrow & {\bf 1} \oplus
[{\bf 7}] \oplus {\bf 7} \oplus 
[{\bf 14}]  \oplus
{\bf 27},  
\label{56}
\eea
and 
the seven Goldstino modes that are absorbed into massive spin
$\frac{3}{2}$ fields (\ref{8}) are identified with septet in
square bracket and the  adjoint representation 
fourteen(${\bf 14}$) in square bracket
corresponds to the massless vector multiplet of the ${\cal N}=1$ theory.
From the further branching rule of $G_2 \rightarrow SU(3)$ by (\ref{g2su3})
one sees that 
the above twenty-seven representation$({\bf 27})$ 
which will be located at Wess-Zumino multiplet  breaks into a
singlet$({\bf 1})$(entering
into the component of ${\cal N}=2$ long massive vector multiplet),
triplet$({\bf 3})$ and 
anti-triplet$({\bf \bar{3}})$(entering into the component of 
${\cal N}=2$ short massive gravitino multiplet), 
sextet$({\bf 6})$ and anti-sextet$({\bf \bar{6}})$(entering 
into the component of ${\cal N}=2$ short
massive hypermultiplet) and octet$(\bf 8)$(entering into the component of 
${\cal N}=2$  massless vector multiplet)
under the $G_2 \rightarrow SU(3)$ breaking.
Moreover, a singlet in (\ref{56})  
which will be located at Wess-Zumino multiplet 
goes to  
the component of ${\cal N}=2$ long massive vector
multiplet while 
the septet in (\ref{56}) 
which will be located at massive gravitino multiplet 
goes to  the component of
${\cal N}=2$ long massive vector multiplet and 
the component of 
${\cal N}=2$ short massive gravitino multiplet.

For spin 1 field, one has the folllowing breaking 
under  $SO(8) 
\rightarrow G_2$ leading to two fundamentals and adjoint
\bea
\mbox{Spin} \; 1: \qquad
{\bf 28} \rightarrow 
{\bf 7} \oplus {\bf 7} \oplus 
[{\bf 14}]
\label{28}
\eea
which
implies that the representation fourteen$({\bf 14})$ in square bracket
corresponds to the massless vector multiplet of the ${\cal N}=1$ theory.
The two septets 
which will be located at massive gravitino multiplet
break into two triplets and
anti-triplets, that enter into the component of 
${\cal N}=2$ short massive gravitino multiplet, a singlet which enters into 
the component of ${\cal N}=2$ massless graviton multiplet and a
singlet that enters into  the component of ${\cal N}=2$ long massive 
vector multiplet
under the further $G_2 \rightarrow SU(3)$ breaking.

For spin zero field, the breaking goes to a sum of 
two singlets, two fundamentals and two symmetric representations
\bea
\mbox{Spin} \; 0: \qquad
{\bf 70}  \rightarrow 
{\bf 1} \oplus
{\bf 1} \oplus [{\bf 7} \oplus 
{\bf 7}]  \oplus
{\bf 27} \oplus {\bf 27}
\label{70}
\eea
under 
 $SO(8) 
\rightarrow G_2$
and 
the fourteen Goldstone bosons modes 
are identified with two septets  in
square bracket. Their quantum numbers are in agreement with those of
massive vectors in (\ref{28}).
The two twenty-seven representations 
which will be located at Wess-Zumino multiplet
break into two singlets(entering
into the component of ${\cal N}=2$ long massive vector multiplet),
triplet and 
anti-triplet(entering into the component of 
${\cal N}=2$ short massive gravitino multiplet), 
two sextets and two anti-sextets(entering into the component 
of ${\cal N}=2$ short
massive hypermultiplet) and two octets(entering into the component of 
${\cal N}=2$  massless vector multiplet)
under the further $G_2 \rightarrow SU(3)$ breaking.
 The remaining triplet and anti-triplet  can be identified
with six Goldstone boson modes under the further 
$G_2 \rightarrow SU(3)$ breaking. 
Moreover, two singlets  in (\ref{70})
which will be located at Wess-Zumino multiplet
enter into 
the component of ${\cal N}=2$ long massive vector
multiplet.
 
Finally, from the breaking
\bea
\mbox{Spin} \; 2: \qquad
{\bf 1} \rightarrow [{\bf 1}]
\label{1}
\eea
under  $SO(8) 
\rightarrow G_2$,
this field
is located at ${\cal N}=1$ massless graviton multiplet.
Of course, under the $SU(3)$, this enters into 
the component of ${\cal N}=2$ massless graviton multiplet. 

We'll reorganize (\ref{8}), (\ref{56}), (\ref{28}), (\ref{70}) 
and (\ref{1})
in the context of supergravity multiplet with corresponding $OSp(1|4)$
quantum numbers in section 4.  
The singlets are placed at Wess-Zumino multiplet and massless graviton
multiplet, septets are located at massive gravitino multiplet,
the adjoints are at massless vector multiplet and 
twenty seven representations sit in another Wess-Zumino multiplet.

\section{An ${\cal N}=1$ supersymmetric membrane flow in three
  dimensional deformed theory }

Let us describe the deformed BL theory by adding 
eight mass parameters $m_1, \cdots, m_8$. 
The theory by \cite{ABJM} is closely related to the BL theory.
Moreover, it is conjectured in \cite{BKKS} that
the $SU(3)_I \times U(1)_Y$-invariant 
$U(2) \times U(2)$ Chern-Simons gauge theory with level 
$k=1$ with the effective 6-th order superpotential is dual to 
the background by two units of four-form flux while the theory at level 
$k=2$ is dual to a ${\bf Z}_2$ orbifold of the background in 
\cite{CPW,AI02-1}.
Recall that the self-dual and anti
self-dual tensors that are invariant under the $G_2$ symmetry 
in ${\cal N}=8$ gauged supergravity
are given by \cite{Warner83,AR99,AP,AW,AW02,AI02,AI02-1,AW03}
\bea
  Y^{+}_{ijkl} & = & 
 [ (\de^{1234}_{ijkl}+\de^{5678}_{ijkl})+
 (\de^{1256}_{ijkl}+ \de^{3478}_{ijkl})+(\de^{3456}_{ijkl}
 +\de^{1278}_{ijkl})] \nonu \\
        &+ & [-(\de^{1357}_{ijkl}
+\de^{2468}_{ijkl})+(\de^{2457}_{ijkl}
     +\de^{1368}_{ijkl})+(\de^{2367}_{ijkl} +\de^{1458}_{ijkl})+
   (\de^{1467}_{ijkl}+\de^{2358}_{ijkl})],
\nonu \\ 
       Y^{-}_{ijkl} & = & i
 [ (\de^{1234}_{ijkl}-\de^{5678}_{ijkl})+
 (\de^{1256}_{ijkl}- \de^{3478}_{ijkl})+(\de^{3456}_{ijkl}
 -\de^{1278}_{ijkl})] \nonu \\
        &+ & i [-(\de^{1357}_{ijkl}
-\de^{2468}_{ijkl})+(\de^{2457}_{ijkl}
     -\de^{1368}_{ijkl})+(\de^{2367}_{ijkl} -\de^{1458}_{ijkl})+
   (\de^{1467}_{ijkl}-\de^{2358}_{ijkl})].
\label{tensor}
\eea 
Turning on the scalar fields proportional to $ Y^{+}_{ijkl}$
yields an $SO(7)^{+}$ invariant vacuum while 
turning on the pseudo-scalar fields proportional to $ Y^{-}_{ijkl}$
yields an $SO(7)^{-}$ invariant vacuum.
Simultaneous turning on both scalar and pseudo-scalar fields leads to
$G_2$-invariant vacuum with ${\cal N}=1$ supersymmetry. 
The choice of the mass parameters  we describe 
corresponds to the self-dual tensor or anti self-dual tensor 
for the indices $5678, 3478, 3456, 2468, 2457, 2367, 2358$, if we shift all the
   indices by adding $2$ to (\ref{tensor}), besides an identity. 
In \cite{HLL} there are three mass parameters
while in \cite{Ahn08} there exist four mass parameters.
Then the fermionic mass terms
from \cite{BL0711} are given by 
\bea
{\cal L}_{f.m.} & = & -\frac{i}{2} h_{ab} \bar{\Psi}^a 
\left(  m_1 \Gamma^{78910}+
  m_2\Gamma^{56910}-
  m_3\Gamma^{5678}+
  m_4\Gamma^{46810} \right. \nonu \\
& + & \left.  m_5\Gamma^{4679} +  m_6\Gamma^{4589} -
  m_7\Gamma^{45710} -  m_8 {\bf 1} \right) \Psi^b.
\label{fermionic}
\eea
The indices $a,b, \cdots$ run over the adjoint of the Lie algebra.
Then the corresponding  
fermionic supersymmetric transformation is given by
\bea
\delta_m \Psi^a  & = & \left(   m_1 \Gamma^{78910}+
  m_2\Gamma^{56910}-
  m_3\Gamma^{5678}+
  m_4\Gamma^{46810} \right. \nonu \\
& + & \left.  m_5\Gamma^{4679} +  m_6\Gamma^{4589} -
  m_7\Gamma^{45710} -  m_8 {\bf 1} \right) X_I^a \Gamma_I \epsilon.
\label{mod1}
\eea
We impose the constraints on the $\epsilon$ parameter that satisfies the 
$\frac{1}{8}$ BPS condition(the number of supersymmetries is two):
\bea
\Gamma^{5678}\epsilon & = & \Gamma^{56910}\epsilon=
\Gamma^{78910}\epsilon = \Gamma^{46810} \epsilon= \Gamma^{34910}
\epsilon
=
\Gamma^{3478}\epsilon= 
\Gamma^{3456}\epsilon \nonu \\
&=&
 -\Gamma^{4679} \epsilon=
-\Gamma^{4589}\epsilon= 
-\Gamma^{45710}\epsilon=
-\epsilon.
\label{cond}
\eea 
The first three conditions in (\ref{cond}) 
provide $\frac{1}{4}$ BPS condition and 
the remaining seven conditions restrict to the $\epsilon$
parameter further and we are left with the right number of
supersymmetry we are dealing with. 

Let us introduce the bosonic mass term which preserves ${\cal N}=1$ 
supersymmetry:
\bea
{\cal L}_{b.m.} = -\frac{1}{2} h_{ab} X_I^a  (m^2)_{IJ} X_J^b.
\label{bosonic}
\eea
Using the supersymmetry variation for $X_I^a$,
$\delta X_I^a = i \bar{\epsilon} \Gamma_I \Psi^a$, and
the supersymmetry variation for $\Psi^a$ by the equation (\ref{mod1}),
the variation for  the bosonic mass term (\ref{bosonic}) plus the
fermionic mass term (\ref{fermionic}) leads to
\bea
\delta {\cal L} & = &  i h_{ab} X_I^a  (m^2)_{IJ} \bar{\Psi}^b \Gamma_J \epsilon  
\nonu \\
&-& i h_{ab} \bar{\Psi}^a  \left(   m_1 \Gamma^{78910}+
  m_2\Gamma^{56910}-
  m_3\Gamma^{5678}+
  m_4\Gamma^{46810} \right. \nonu \\ 
& + &   \left. m_5\Gamma^{4679} +  m_6\Gamma^{4589} -
  m_7\Gamma^{45710} -  m_8 {\bf 1} \right)^2
X_I^b \Gamma_I \epsilon. 
\label{van}
\eea
In order to vanish this, the bosonic mass term $(m^2)_{IJ} \Gamma_J$
should take the form
\bea
(m^2)_{IJ}\Gamma^J \rightarrow
&& (m_1+m_2-m_3+m_4 -m_5 -m_6 +m_7 -m_8)^2 \Gamma_3
\nonu \\
&+& (m_1+m_2-m_3-m_4 +m_5 +m_6 -m_7 -m_8)^2 \Gamma_4
\nonu \\
&+& (m_1-m_2+m_3+m_4 -m_5 +m_6 -m_7 -m_8)^2 \Gamma_5 \nonu \\
& + & (m_1-m_2+m_3-m_4 +m_5 -m_6 +m_7 -m_8)^2 \Gamma_6
\nonu \\
 &+ & (m_1-m_2-m_3-m_4 -m_5 +m_6 +m_7 +m_8)^2 \Gamma_7
\nonu \\
&+ & (m_1-m_2-m_3+m_4 +m_5 -m_6 -m_7 +m_8)^2 \Gamma_8
\nonu \\
& + & (m_1+m_2+m_3-m_4 -m_5 -m_6 -m_7 +m_8)^2 \Gamma_9\nonu \\
&+ &
(m_1+m_2+m_3+m_4 +m_5 +m_6 +m_7 +m_8)^2 \Gamma_{10}
\label{massdia}
\eea
 by
computing the mass terms for second and third lines 
of (\ref{van}) explicitly \footnote{The
  relevant terms 
  become $m_1^2 +m_2^2 +m_3^2 +m_4^2+m_5^2 +m_6^2+m_7^2+m_8^2 +2(-m_1
  m_2-m_4 m_7-m_5 m_6 +m_3 m_8)
  \Gamma^{5678}
+ 2(m_1
  m_3+m_4 m_6+m_5 m_7 -m_2 m_8)\Gamma^{56910}
+2(m_1
  m_4+m_2 m_7-m_3 m_6 -m_5 m_8)\Gamma^{4679}+
2(m_1
  m_5+m_2 m_6+m_3 m_7 -m_4 m_8)\Gamma^{46810} + 2(-m_1
  m_6-m_2 m_5-m_3 m_4 +m_7 m_8)\Gamma^{45710} +2(m_1
  m_7+m_2 m_4+m_3 m_5 -m_6 m_8)\Gamma^{4589}
 +
2(m_2
  m_3+m_4 m_5+m_6 m_7 -m_1 m_8) \Gamma^{78910} 
 $ explicitly. We also used the conditions (\ref{cond}).}.
When all the mass parameters are equal 
\bea
m_1 =m_2 =m_3=m_4=m_5=m_6=m_7=m_8=m,
\nonu
\eea
then the diagonal bosonic mass term in (\ref{massdia}) has nonzero
component only for $1010$ and other components($33, 44, 55,
66, 77, 88$ and $99$) are vanishing. This resembles the structure of
$A_1^{IJ}$ tensor of $AdS_4$ supergravity where the $A_1^{IJ}$ 
tensor has two
distinct eigenvalues with degeneracies $7$ and $1$ respectively 
\cite{dNW,AI02}. The
degeneracy $1$ is related to the ${\cal N}=1$ supersymmetry.  
Then one obtains the bosonic mass term which 
appears in (\ref{bosonic})
\bea
(m^2)_{IJ} = \diag(0,0,0,0,0,0,0,64m^2).
\label{massmass}
\eea

Let us introduce the eight ${\cal N}=1$ superfields
as follows:
\bea
\Phi_1  & = &  X_3 + \cdots, 
\qquad \Phi_2 =X_4  + \cdots, 
\qquad \Phi_3 =X_5 + \cdots, 
\nonu \\
\Phi_4  & = &  X_6 + \cdots,  
\qquad \Phi_5 =X_7  + \cdots,
\qquad \Phi_6 =X_8 + \cdots, 
\nonu \\
\Phi_7  & = &  X_9 + \cdots,  
\qquad \Phi_8 =X_{10}  + \cdots,
\label{n1}
\eea
where we do not include the ${\cal N}=1$ fermionic fields.
The $\Phi_1, \cdots, \Phi_7$ constitute a fundamental ${\bf 7}$ representation
of $G_2$ while $\Phi_8$ is a singlet ${\bf 1}$ of $G_2$.
The potential in the BL theory \cite{BL0711} is given by
\bea
\frac{1}{3\kappa^2} h_{ab} f_{cde}^{\;\;\;\;a} 
X_I^c X_J^d X_K^e f_{fgh}^{\;\;\;\;b} X_I^f X_J^g X_K^h
\nonu
\eea 
where $\kappa$ is a Chern-Simons coefficient.
In terms of ${\cal N}=1$ superfields, this contains the following expressions
\bea
\frac{1}{\kappa^2} h_{ab} f_{cde}^{\;\;\;\;a} f_{fgh}^{\;\;\;\;b}
\left( \Phi_1^c \Phi_2^d \Phi_3^e  \Phi_1^f
\Phi_2^g 
\Phi_3^h + \mbox{other terms} 
\right)
\nonu
\eea
by using the relation (\ref{n1}) between the component fields and superfields.
This provides the superpotential and it is given by \cite{MP} with (\ref{tensor})
\bea
\frac{1}{\kappa} f_{abcd}
Y^{+ijkl} \Tr \Phi_i^a \Phi_j^b \Phi_k^c \Phi_l^d.
\label{inv}
\eea
This form has manifest $SO(7)^{+}$ global symmetry and by 
fixing the coefficients of ${\cal N}=1$ superspace action \cite{MP}
this global symmetry is enhanced to $SO(8)$ symmetry with maximal supersymmetry.
In ${\cal N}=1$ language, the superpotential consisting of the mass
term (\ref{massmass}) 
and quartic term (\ref{inv}) where we absorbed the $\kappa$ into the
structure constant is given by
\bea
W  & = & \frac{1}{2} M \Tr \Phi_8^2 
+ \left( \Tr 
  \Phi_1 \Phi_2 \Phi_7 \Phi_8  + \mbox{other 6-terms with $\Phi_8$} \right) 
\nonu \\
& + & 
\left(  
\Tr \Phi_3 \Phi_4 \Phi_5 \Phi_6 + \mbox{other 6-terms without $\Phi_8$} 
 \right).
\nonu
\eea
The fourteen terms except the first term are 
the superpotential required by 
${\cal N}=8$ supersymmetry and 
the first term breaks ${\cal N}=8$ down to ${\cal N}=1$.
The theory has matter multiplet in seven flavors $\Phi_1, \Phi_2, \cdots,
\Phi_7$
transforming in the adjoint with 
flavor symmetry under which the matter multiplet
forms a septet$({\bf 7})$ of the ${\cal N}=1$ theory.
Therefore, we turn on the mass perturbation in the UV and flow to the IR.
This maps to turning on certain fields in the $AdS_4$
supergravity where they approach to zero in the UV($r
\rightarrow \infty$) and develop a nontrivial profile as a function of
$r$ as one goes to the
IR($r \rightarrow -\infty$).
We can integrate out the massive scalar $\Phi_8$ with adjoint index 
at a low enough scale
and this results in the 6-th order superpotential $\Tr (Y^{+ijk8} \Phi_i 
\Phi_j \Phi_k)^2 + \Tr \epsilon_{ijklmnp} Y^{+ijk8}(\Phi_l \Phi_m \Phi_n \Phi_p)$.  
The scale dimensions of eight superfields $\Phi_i(i=1, 2, \cdots, 8)$ 
are $\Delta_i = \frac{1}{4}$ at the UV. 
This is because the sum of $\Delta_i$ is equal to the canonical dimension
of the superpotential which is $3-1=2$ \cite{Strassler98}.
By symmetry, one arrives at $\Delta_i=\frac{1}{4}$. 

Thus we have found ${\cal N}=1$ superconformal Chern-Simons theories
with global $G_2$ symmetry and $k=1$ Chern-Simons gauge theories with 
$G_2$-invariant superpotential deformation are dual to the holographic
RG flows in \cite{AI02}.
We expect that 
$G_2$-invariant $U(N) \times U(N)$ Chern-Simons gauge theory 
for $N > 2$ with $k=1,2$ where there exists an enhancement of
${\cal N}=8$ supersymmetry \cite{ABJM,BKKS}  
is dual to the background of \cite{AI02} with $N$ unit of flux.
In next section, the gauge invariant composites  
in the superconformal field theory at
the IR in three dimensions 
are mapped to the corresponding supergravity bulk fields 
in four dimensions.

\section{The $OSp(1|4)$ spectrum and operator map between bulk and
  boundary theories }

A further detailed correspondence between fields of
$AdS_4$ supergravity in four dimensions and composite operators of the 
IR field theory in three dimensions is described in this section.

The even subalgebra of the superalgebra $OSp(1|4)$
is $Sp(4,R)\simeq SO(3,2)$ that is the isometry algebra of $AdS_4$
\cite{Heidenreich}.
The maximally compact subalgebra is then 
$SO(2)_E \times SO(3)_S $
where the generator of $SO(2)_E$ is the hamiltonian of the system
and its eigenvalues $E$ are the energy levels of states for the
system,
the group $SO(3)_S$ is the rotation group and its representation $s$ 
describes the spin states of the system.
A supermultiplet, a unitary irreducible representations(UIR) of the
superalgebra
$OSp(1|4)$, consists of a finite number of UIR of the even subalgebra
and a particle state is characterized by a spin $s$ and energy $E$. 

Let us classify the supergravity multiplet, which is invariant under 
$G_2$, we explained in section 2 
and describe them in the three dimensional 
boundary theory.

$\bullet$ Wess-Zumino multiplet

The conformal dimension $\Delta$, which is irrational and unprotected, 
is given by $\Delta=E_0 > \frac{1}{2}$.
Let us denote $S(x, \theta)$, that is a 
scalar superfield, by the corresponding boundary operator 
in boundary gauge theory.
This scalar field has a dimension $ \frac{5}{6}(6+\sqrt{3})$ in the IR.
We'll come back this issue at the end of this section.
The corresponding $OSp(1|4)$ representations and corresponding
${\cal N}=1$ superfield in three dimensions are listed in Table 2.

\begin{table} 
\begin{center}
\begin{tabular}{|c|c|c|c|} \hline
Boundary Operator & Energy & Spin $0$  & Spin $\frac{1}{2}$ 
 \\ \hline
$S = \left( \sum_{i=1}^{7} \Phi_i^2 
\right)^{\frac{6}{5}}$ & $E_0$ & ${\bf 1}$ &     \\
& $E_0+\frac{1}{2}$ & & 
${\bf 1}$  
   \\
& $E_0+1$ &  ${\bf 1} $ &    \\
\hline 
\end{tabular} 
\end{center}
\caption{\sl 
The $OSp(1|4)$ representations(energy, spin) 
and $G_2$ representations singlets
in the supergravity mass spectrum for
Wess-Zumino multiplet at
the ${\cal N}=1$ critical point and the corresponding ${\cal N}=1$
superfield $S$ in the boundary gauge theory.}
\end{table} 

$\bullet$ Wess-Zumino multiplet

The conformal dimension $\Delta$ for the lowest
component of this multiplet is given by $\Delta =E_0 > \frac{1}{2}$.
The $AdS_4$ supergravity multiplet corresponds to the 
scalar superfield $\Phi(x, \theta)$. 
That is, in the $\theta$ expansion, there are three component
fields in the bulk.  
Then the bilinear of the seven $\Phi_i$ superfields by symmetrizing the two $G_2$
indices present in them 
provides a symmetric representation of $G_2$, ${\bf 27 }$,
corresponding to $\Tr \Phi_{(i} \Phi_{j)}$.
Note that from the tensor product between two ${\bf 7}$'s
of $SO(7)$, one writes down 
${\bf 7} \times {\bf 7} = {\bf 1}_s \oplus {\bf 21}_a \oplus
{\bf 27}_s$. Using the branching rule of (\ref{so7g2}), this leads to 
${\bf 1} \oplus {\bf 7} \oplus {\bf 14} \oplus {\bf 27}_s$ under the
breaking $SO(7)$ into $G_2$.
Therefore, one obtains the symmetric 
representation ${\bf 27}$ of $G_2$.
Group theoretically, the structure of this 
Wess-Zumino multiplet and ${\cal N}=2$ short massive hypermultiplet
in \cite{Ahn08} looks similar. The symmetric representations ${\bf 6}$
and ${\bf \bar{6} }$ of $SU(3)$ originate from only this symmetric
representation ${\bf 27}$ of $G_2$ when we look at the branching rule
by (\ref{g2su3}). As we observed in section 2, 
through the supersymmetry enhancement from ${\cal N}=1$ to ${\cal
N}=2$,
some other components among the repesentation ${\bf 27}$ of $G_2$
distribute into other ${\cal N}=2$ multiplets  while
the symmetric representations ${\bf 6}$
and ${\bf \bar{6} }$ of $SU(3)$ remain and constitute 
${\cal N}=2$ short massive hypermultiplet.
The corresponding $OSp(1|4)$ representations and corresponding
superfield are listed in Table 3.

\begin{table} 
\begin{center}
\begin{tabular}{|c|c|c|c|} \hline
Boundary Operator & Energy & Spin $0$  & Spin $\frac{1}{2}$ 
 \\ \hline
$\Tr \Phi_{(i} \Phi_{j)}$& 
$E_0$ & ${\bf 27}$ &     \\
 & $E_0+ \frac{1}{2}$ & & 
${\bf 27}$     \\
& $E_0+1$ &  ${\bf 27} $ &   \\
\hline 
\end{tabular} 
\end{center}
\caption{\sl 
The $OSp(1|4)$ representations(energy, spin) 
and $G_2$ symmetric representations in the supergravity mass spectrum for
Wess-Zumino multiplet at
the ${\cal N}=1$ critical point and the corresponding ${\cal N}=1$
superfield in the boundary gauge theory.}
\end{table} 

$\bullet$ Massive gravitino multiplet

The conformal dimension $\Delta$ is given by $\Delta = E_0 > 2$.
This  corresponds to 
spinorial superfield $\Phi_{\alpha\beta}(x, \theta)$.  
In the $\theta$ expansion, the component
fields in the bulk are located with appropriate quantum numbers. 
Then one can identify $\Tr D_{\alpha} W_{\beta} \Phi_j$ with ${\bf 7}$ of $G_2$.
Although the group structure between 
this 
massive gravitino multiplet and 
${\cal N}=2$ short massive gravitino 
multiplet
is different
from each other, 
they have both fundamental representations and 
one obtains this 
multiplet when one takes ${\cal N}=1$ superderivative 
$D_{\alpha}$ on ${\cal N}=2$ short massive gravitino 
multiplet in \cite{Ahn08}.
Some of  the fundamental and anti-fundamental 
representations ${\bf 3}$
and ${\bf \bar{3} }$ of $SU(3)$ in ${\cal N}=2$ theory 
originate from this fundamental
representation ${\bf 7}$ of $G_2$ when we look at the branching rule
by (\ref{g2su3}) and some of them come from 
others ${\bf 14}$ and ${\bf 27}$ of $G_2$. 
As we observed in section 2, 
from ${\cal N}=1$ to ${\cal
N}=2$,
the singlets among the representation ${\bf 7}$ of $G_2$
enter 
into ${\cal N}=2$ massless graviton multiplet  or 
long massive vector multiplet.
The corresponding $OSp(1|4)$ representations and corresponding
superfield are listed in Table 4.

\begin{table} 
\begin{center}
\begin{tabular}{|c|c|c|c|c|} \hline
Boundary Operator                    & Energy             
& Spin $\frac{1}{2}$ & Spin $1$ & Spin $\frac{3}{2}$ \\ \hline 
$\Tr D_{\alpha} W_{\beta} \Phi_j$ & $E_0$ &  & ${\bf 7}$ &   \\
 & $E_0 +\frac{1}{2} $  & ${\bf 7} $ & & ${\bf 7}$  \\
& $E_0+1 $ &   
& ${\bf 7}$  &   \\
\hline 
\end{tabular} 
\end{center}
\caption{\sl 
The $OSp(1|4)$ representations(energy, spin) 
and  $G_2$ fundamental representations in the supergravity mass spectrum for
massive gravitino multiplet at
the ${\cal N}=1$ critical point and the corresponding ${\cal N}=1$
superfield in the boundary gauge theory.}
\end{table} 

$\bullet$ ${\cal N}=1$ massless graviton multiplet

The bulk field $\Phi_{\alpha\beta\gamma}(x,\theta)$ 
can be identified with $D^{\alpha} T^{\beta\gamma}(x, \theta)$
where  $T^{\beta \gamma}(x, \theta)$ is the  
stress energy tensor superfield. 
In components, 
the $\theta$ expansion of this superfield has
the ${\cal N}=1$ supercurrent and the stress energy tensor.
Also in this case,
one obtains this 
multiplet when one takes ${\cal N}=1$ superderivative 
$D^{\alpha}$ on ${\cal N}=2$ massless graviton 
multiplet in \cite{Ahn08}.
The conformal dimension $\Delta=\frac{5}{2}$.
The structure of this 
massless graviton multiplet and ${\cal N}=2$ massless graviton
multiplet
 looks similar to each other. 
Some of  the singlet
representation ${\bf 1}$
of $SU(3)$ in ${\cal N}=2$ theory originate from this singlet
representation ${\bf 1}$ of $G_2$ when we look at the branching rule
by (\ref{g2su3}) and some of them come from 
${\bf 7}$ and ${\bf 27}$ of $G_2$. 
The corresponding $OSp(1|4)$ representations and corresponding
superfield are listed in Table 5.

$\bullet$ ${\cal N}=1$ massless vector multiplet

This conserved vector current is given by 
a scalar superfield $D_{\alpha} J^A(x, \theta)$. 
This transforms in the
adjoint representation of $G_2$ flavor group. The boundary object
is given by 
$\Tr D_{\alpha} \Phi T^A \Phi$ where 
the generator $T^A$ is   $N \times
N$ matrix  with $A=1, 2, \cdots, N^2-1$ for general $N$. 
One obtains also this 
multiplet when one takes ${\cal N}=1$ superderivative 
$D_{\alpha}$ on ${\cal N}=2$ massless vector 
multiplet in \cite{Ahn08} although the group structure is different
from each other but they have both adjoint representations.
The conformal dimension $\Delta=\frac{3}{2}$.
By taking a tensor product between two ${\bf 7}$'s, one
gets this adjoint ${\bf 14}$ of $G_2$ representation as in Wess-Zumino
multiplet above.
The structure of this 
massless vector multiplet and ${\cal N}=2$ massless vector 
multiplet resembles each other. 
Some of  the adjoint
representation ${\bf 8}$
of $SU(3)$ in ${\cal N}=2$ theory originate from this adjoint
representation ${\bf 14}$ of $G_2$ when we look at the branching rule
by (\ref{g2su3}) and some of them come from 
${\bf 27}$ of $G_2$. 
As we observed in section 2, 
from ${\cal N}=1$ to ${\cal
N}=2$,
triplets and anti-triplets among the representation ${\bf 14}$ of $G_2$
enter 
into ${\cal N}=2$ short massive gravitino multiplet.
The corresponding $OSp(1|4)$ representations and corresponding
superfield are listed in Table 5 also.

\begin{table} 
\begin{center}
\begin{tabular}{|c|c|c|c|c|c|} \hline
Boundary Operator & Energy & Spin $\frac{1}{2}$ & Spin $1$
 & Spin $\frac{3}{2}$ & Spin $2$ \\ \hline 
$\Tr  D_{\alpha} \Phi T^A \Phi $ 
& $E_0 = \frac{3}{2}$ & $[{\bf 14}]$ &  & & \\
& $E_0+\frac{1}{2} = 2$ &  
& $[{\bf 14}]$  &  & \\
\hline
$D^{\alpha} T^{\beta\gamma}$  
 & $E_0 =\frac{5}{2}$ 
&  & & $[{\bf 1}]$ & \\
& $E_0 +\frac{1}{2} = 3$  & & & & $[{\bf 1}]$ \\
\hline
\end{tabular} 
\end{center}
\caption{\sl 
The $OSp(1|4)$ representations(energy, spin) 
and  $G_2$ representations(adjoints and singlets) 
in the supergravity mass spectrum for
massless vector and graviton multiplets at
the ${\cal N}=1$ critical point and the corresponding ${\cal N}=1$
superfields in the boundary gauge theory.}
\end{table} 

The 11-dimensional metric with warped product ansatz
is given by \cite{dNW,dN87,AI02,AI02-1}
\bea
ds_{11}^2 = 
ds_4^2 + ds_7^2 =
\Delta(x,y)^{-1} \,g_{\mu \nu}(x) \,d x^{\mu} d x^{\nu}
+ G_{mn}(x,y) \,dy^m dy^n, 
\nonu
\eea 
where $\mu, \nu =1, 2, \cdots, 4$ and $m, n=1, 2, \cdots, 7$.
The 4-dimensional metric which has a 3-dimensional Poincare invariance
takes the form
$
g_{\mu \nu}(x) \,d x^{\mu} d x^{\nu} = e^{2A(r)} \,\eta_{\mu' \nu'}\,
d x^{\mu'} d x^{\nu'} + dr^2$, 
where $\eta_{\mu' \nu'}=(-,+,+)$ and $r=x^4$ is the coordinate
transverse to the domain wall as in section 2 
and the scale factor $A(r)$ behaves
linearly in $r$ at UV and IR regions. 
The metric formula by \cite{dNW} generates the 7-dimensional metric
from the two input data of $AdS_4$ vacuum expectation values
for scalar and pseudo-scalar fields $(\la, \al)$.
Let us introduce the redefined fields \cite{AI02}
\begin{eqnarray}
a \equiv \cosh\!\left(\frac{\lambda}{\sqrt{2}}\right)
+\cos\alpha\,\sinh\!\left(\frac{\lambda}{\sqrt{2}}\right),\qquad
b \equiv \cosh\!\left(\frac{\lambda}{\sqrt{2}}\right)
-\cos\alpha\,\sinh\!\left(\frac{\lambda}{\sqrt{2}}\right).
\nonu
\end{eqnarray}
We recall that 
the two input data of $(a, b)$ are 
\bea
a=1, \qquad b=1
\label{abso8}
\eea
for the $SO(8)$-invariant UV critical point whereas
\bea
a = \sqrt{\frac{6 \sqrt{3}}{5}}, \qquad b = \sqrt{\frac{2\sqrt{3}}{5}}
\label{abvalues}
\eea
for the $G_2$-invariant IR critical point \footnote{The scalar
  potential (\ref{pot}) can be rewritten as $V(a,b)=\frac{1}{8} a^2\left(
    a^5-28a^2 b + 14a^3 b^2 -84 b^3 + 49 ab^4\right)$.}.

From the standard metric of a 7-dimensional ellipsoid, the diagonal 
$8 \times 8$ matrix $Q_{AB}$ is given by 
$
Q_{AB}={\rm diag}\left(b^2, b^2, b^2,  b^2, b^2, b^2, b^2, a^2\right)
$  \cite{AI02,AI02-1}
and the 7-dimensional ellipsoidal metric 
$ds_{EL(7)}^2=dX^A Q^{-1}_{AB}\,dX^B$ 
where $X^A$ is a coordinate for $R^8$ and $A,B=1, 2, \cdots, 8$ 
arises via 
\begin{equation}
ds_7^2 =G_{mn}(x, y) \,dy^m dy^n
=\sqrt{\Delta \,a}\; L^2 \left(\,
a^{-2}\,\xi^2\,d\theta^2 +\sin^2 \theta\,d\Omega_6^2 
\,\right)
\label{7metric}
\end{equation}
where the quadratic form $\xi^2$ is given by 
\cite{dNW,AI02,AI02-1}
\bea
\xi^2 =a^2 \cos^2 \theta + b^2 \sin^2 \theta
\label{xi}
\eea
and the warped factor $\Delta$ is given by
\bea
\Delta=a^{-1}\,\xi^{-{4 \over 3}}.
\label{Delta}
\eea
The metric  $d\Omega_6^2$ on ${\bf S}^6 \simeq G_2/SU(3)$ in (\ref{7metric})  
preserves 
the Fubini-Study metric on ${\bf CP}^2$ \cite{GW,AI02-1}.
Note that the corresponding base 6-sphere for $SU(3)_I \times
U(1)_R$-invariant  sector \cite{NW,Ahn08} 
is given by ${\bf CP}^3$ which is the homogeneous 
space $SU(4)/[SU(3) \times U(1)]$ characterized by 
the Kahler form $J$ \cite{PW,AI02-1}. 

As in \cite{CPW}, let us go to the $SL(8,R)$ basis and introduce 
the rotated vielbeins
\bea
U^{ij}_{\,\,\,\,IJ} & = & u^{ij}_{\,\,\, ab} (\Gamma_{IJ})^{ab}, 
\qquad
V^{ijIJ} = v^{ijab} (\Gamma_{IJ})^{ab}
\nonu \\
U_{ij}^{\,\,\,\,IJ} & = & u_{ij}^{\,\,\, ab} (\Gamma_{IJ})^{ab}, 
\qquad
V_{ijIJ} = v_{ijab} (\Gamma_{IJ})^{ab}
\nonu
\eea
where 28-beins and gamma matrices are the same 
as those in \cite{AW}.
Now let us define
\bea
A_{ijIJ} & = & 
\frac{1}{\sqrt{2}} \left( U_{ij}^{\,\,\,\,IJ} + V_{ijIJ}
\right), 
\qquad
B_{ij}^{\,\,\,IJ} =
\frac{1}{\sqrt{2}} \left( U_{ij}^{\,\,\,\,IJ} - V_{ijIJ} \right),
\nonu \\
C^{ij}_{\,\,\,IJ} & = & \frac{1}{\sqrt{2}} \left(U^{ij}_{\,\,\,\,IJ} +
V^{ijIJ}  \right), \qquad 
D^{ijIJ} =\frac{1}{\sqrt{2}} \left(-U^{ij}_{\,\,\,\,IJ} +
V^{ijIJ}  \right).
\nonu
\eea
Then ``geometric'' $T$-tensor can be 
written as
\bea
\widetilde{T}_l^{\,kij} = \frac{1}{168\sqrt{2}} C^{ij}_{\,\,LM} 
\left( A_{lmJK} D^{kmKI} \, \delta^L_{\,I} \, x_M x_J -B_{lm}^{\,\,\,JK} 
C^{km}_{\,\,\,KI} \, \delta^M_{\,J} \, x_L x_I \right)
\label{ttilde}
\eea
and furthermore the ``geometric'' $A_1$-tensor
is given by
\bea
\widetilde{A}_1^{\,ij} = \widetilde{T}_m^{\,\,imj}.
\label{a1tilde}
\eea
The idea of \cite{KW} is to replace $\delta^{IJ}$ in the original
$T$-tensor 
with $x^I x^J$ but
$\delta^I_J$ remains unchanged, as in (\ref{ttilde}).

Now the $88$ component of  $\widetilde{A}_1^{\,ij}$,
$\widetilde{A}_1^{\,88}$,
provides the ``geometric'' superpotential in terms of $a, b$ and $\theta$  
and from the complete expressions in appendix B (\ref{a1}), one gets 
\bea
W_{gs} \equiv  |\widetilde{A}_1^{\,88}|^2 = 
a^{\frac{3}{2}}\sqrt{(a^2 \cos^2 \theta + b^2 \sin^2
  \theta)^2 
-16(ab-1)\sin^2 \theta \cos^2 \theta}.
\label{wgs}
\eea
This superpotential is different from $W_{AI}$ in \cite{AI02}, in general.
More explicitly, one obtains
\footnote{
The superpotential $W$ is the same as (\ref{W})
$
W= \frac{1}{8} \sqrt{a^3\left[\left( a^2 + 7 b^2 
\right)^2 -112 \left(ab -1\right) \right]}
$
and the superpotential by \cite{AI02}
is
$
W_{AI} = \frac{1}{16 W} a^3 \left[ \left(48(1-a b) +(a^2 -b^2) (a^2 +
    7 b^2 )\right) \cos^2 \theta +
  8\left(1- a b \right) + b^2 \left( a^2 + 7 b^2\right) \right]$.}
\bea
W_{gs}^2 = 4 W_{AI}^2 + \frac{1}{4W^2} a^6 \left(-2 + a b \right)^2
\left(-1+a b \right) \left( 3 + 4 \cos 2\theta \right)^2.
\nonu
\eea
In particular, when $\theta = \cos^{-1} \frac{1}{\sqrt{8}}$, 
one obtains $W_{gs } = W = W_{AI}$.
Although there are two different solutions for the superpotential,
$W_{gs}$
and $W_{AI}$, in 11-dimensions, there exists the same superpotential 
$W$ in 4-dimensions.
For $SO(8)$ maximal ${\cal N}=8$ supersymmetric critical point with 
(\ref{abso8}),
it is easy to check $W_{gs}=2W_{AI}=W=1$. 

Performing the M2-brane probe analysis \cite{JLP00,KW,JLP01},
one can compute the effective Lagrangian for the probe moving 
at a small velocity transverse to its world-volume.
If the potential vanishes, 
then the kinetic term gives us to a metric on the corresponding 
moduli space. By combining some part of determinant for the
induced metric for M2-brane world-volume with 3-form potential,
one reads off the corresponding potential. 
Then the potential has the factor
\bea
\Delta^{-\frac{3}{2}} -W_{gs} = a^{\frac{3}{2}}\left( a^2 \cos^2
  \theta + b^2 \sin^2 \theta\right)\left[\, 1- 
\sqrt{1-\frac{16 \left( a b -1\right) \sin^2 \theta \cos^2 \theta}
{\left( a^2 \cos^2
  \theta + b^2 \sin^2 \theta\right)^2}} \, \right]
\nonu
\eea
where we substituted (\ref{Delta}), (\ref{xi}) and (\ref{wgs}).
This potential vanishes for $\theta=\frac{\pi}{2}$. Of course, there
exists trivial solution for $\theta=0$ where the metric becomes zero.
On this subspace, the metric on the 7-dimensional moduli space
transverse to the M2-branes is given by
\bea
ds^2|_{\rm{moduli}} = 
\sqrt{a}\, L^2 \,e^A\, d \Omega_6^2 + e^A \,a^{\frac{3}{2}}\, b^2\, dr^2
\label{modulimetric}
\eea
where we used the fact that for $\theta=0$ we have simple relations
from (\ref{xi}) and (\ref{Delta}):
$
\xi^2 = b^2$ and $ \Delta = a^{-1} b^{-\frac{4}{3}}$.

As we approach the IR critical point, we can introduce a new radial
coordinate $ u \simeq e^{\frac{1}{2} A(r)}$ with 
$\frac{d u }{d r}    = 
\frac{1}{L} \sqrt{\frac{6}{5}} \sqrt{a}\, b\, u$
to obtain the asymptotic form for the metric by 
inserting (\ref{abvalues}), (\ref{flow}) and the IR critical value of $W$
in Table 1 into (\ref{modulimetric})
\bea
ds^2|_{\rm{moduli}} = \frac{5^{\frac{3}{4}} 3^{\frac{1}{8}}}
{6^{\frac{3}{4}}} L^2 \left( du^2 + 
\frac{6}{5} u^2 \, d \Omega_6^2 \right). 
\label{mod}
\eea 
Also at the IR critical point, one can see that
\bea
\frac{d  }{d r}  \left( e^{A(r)} \sqrt{a} \right)|_{\rm{IR}}  = 
\frac{2}{L}\, \sqrt{\frac{6}{5}}\, a\, b\, e^{A(r)}|_{\rm{IR}}
\nonu
\eea
by using the supersymmetric flow equations (\ref{flow})  
we introduced in section 2.
The mass spectrum for the $\frac{\sqrt{7}}{2} \la$ around $G_2$ fixed
point was computed in \cite{AR99} and it is 
$
\frac{5}{6} \left( 6 + \sqrt{3}\right)$.
At the IR end of the flow, $A(r) \sim \frac{2\sqrt{\frac{36
      \sqrt{2}\,3^{1/4}}
{25 \sqrt{5}}}}{L} r$
with $g \equiv \frac{\sqrt{2}}{L}$ from the solution (\ref{flow})
for $A(r)$ and $W= \sqrt{\frac{36 \sqrt{2}\,3^{1/4}}{25 \sqrt{5}}}$ 
from Table 1. Moreover,
$u \sim
e^{\frac{\sqrt{\frac{36
      \sqrt{2}\,3^{1/4}}
{25 \sqrt{5}}} }{ L}r}  \sim e^{\frac{A(r)}{2}}
$ 
above. 
Then $S$ becomes $S =(\Phi_1^2  +\cdots \Phi_7^2 )^
{\frac{6}{5}}$ in the boundary theory. The power $\frac{6}{5}$ comes
from the factor in the metric (\ref{mod}) of the moduli.
Obviously, from the tensor product between ${\bf 7}$ and ${\bf 7}$
of $G_2$ representation, one gets a singlet ${\bf 1}$ as before. 
For the superfield $S(x, \theta)$, the 
action looks like $\int d^3 x d^2 \theta 
S(x, \theta)$. The component content of this action 
can be worked out straightforwardly using the projection technique. 
This implies that the highest component field in $\theta$-expansion,
the last element in Table 2, has a conformal dimension  $6+ \frac{5}{6} 
\sqrt{3}$ in the IR as before. 

We have presented the gauge invariant 
combinations of the massless superfields 
of the gauge theory whose $G_2$ 
quantum numbers exactly match the four multiplets in
Tables $3,4, 5$
observed in the supergravity.  There exists one additional Wess-Zumino
multiplet in Table $2$ which completes the picture. 

\section{
Conclusions and outlook }

By analyzing the mass-deformed Bagger-Lambert theory(or  
the mass-deformed $U(2) \times
U(2)$ Chern-Simons gauge theory with level $k=1$ or $2$), preserving 
$G_2$ symmetry, with the addition of mass term
for one of the eight adjoint superfields, 
one identifies an ${\cal N}=1$ supersymmetric membrane flow in three
dimensional deformed theory with 
the holographic ${\cal N}=1$ 
supersymmetric RG flow in four dimensions. 
Therefore, the ${\cal N}=8$ gauged supergravity critical point
is indeed the holographic dual of the mass-deformed ${\cal N}=8$ BL
theory(or 
the mass-deformed $U(2) \times
U(2)$ Chern-Simons gauge theory with level $k=1$ or $2$).
So far, we have focused on the particular mass deformation 
(\ref{fermionic}) preserving $G_2$ symmetry. 
As we mentioned in introducation, there are three more
nonsupersymmetric critical points, $SO(7)^{+}, 
SO(7)^{-}$ and $SU(4)^{-}$.
It would be interesting 
to find out all the possible cases for the mass deformations
and see how they appear in the $AdS_4 \times {\bf X}^7$ background in
the context of  $U(N) \times
U(N)$ Chern-Simons gauge theory. 

\vspace{.7cm}
\centerline{\bf Acknowledgments}

I would like to thank K. Hosomichi   for discussion on BL theory 
and T. Itoh, K. Woo and J. Hwang for earlier 
related works on gauged supergravity. 
This work was supported by grant No.
R01-2006-000-10965-0 from the Basic Research Program of the Korea
Science \& Engineering Foundation.  
I would like to thank the participants and the 
organizers of the mini workshop
on ``Chern-Simons theories for M2-branes at CQUeST'' during 
6/26-6/27, 2008 for discussions. 

\appendix

\renewcommand{\thesection}{\large \bf \mbox{Appendix~}\Alph{section}}
\renewcommand{\theequation}{\Alph{section}\mbox{.}\arabic{equation}}

\section{Branching rules}

In this appendix we list some useful branching rules with the help of
\cite{ps}.
In order to obtain the branching rule $SO(8) \rightarrow G_2$, 
we need to consider the following two branching rules,
$SO(8) \rightarrow SO(7)$ and $SO(7) \rightarrow G_2$.
The former is given by 

$\bullet$ $SO(8) \rightarrow SO(7)$  branching rule
\bea
{\bf 1} & \rightarrow & {\bf 1}, \nonu \\
{\bf 8}_v & \rightarrow & {\bf 8}, \nonu \\
{\bf 8}_s & \rightarrow & {\bf 1} \oplus {\bf 7}, \nonu \\
{\bf 8}_c  & \rightarrow & {\bf 8},
\nonu \\
{\bf 28}  & \rightarrow & 
{\bf 7} \oplus {\bf 21}, \nonu \\
{\bf 35}_v & \rightarrow & {\bf 35}, \nonu \\
{\bf 35}_s & \rightarrow & {\bf 1} \oplus {\bf 7} \oplus {\bf 27},
\nonu \\
{\bf 35}_c  & \rightarrow & {\bf 35}, \nonu \\
{\bf 56}_v & \rightarrow & {\bf 8} \oplus {\bf 48}, \nonu \\
{\bf 56}_s  & \rightarrow & {\bf 21} \oplus {\bf 35}, \nonu \\
{\bf 56}_c & \rightarrow & {\bf 8} \oplus {\bf 48}
\label{so8so7}
\eea
and 
the latter is given by

$\bullet$ $SO(7) \rightarrow G_2$  branching rule
\bea
{\bf 1} & \rightarrow & {\bf 1}, \nonu \\
{\bf 7} & \rightarrow & {\bf 7}, \nonu \\
{\bf 8} & \rightarrow & {\bf 1} \oplus {\bf 7}, \nonu \\
{\bf 21} & \rightarrow & {\bf 7} \oplus {\bf 14}, \nonu \\
{\bf 27} & \rightarrow & {\bf 27} \nonu \\
{\bf 35} & \rightarrow & {\bf 1} \oplus {\bf 7} \oplus {\bf 27}, \nonu \\ 
{\bf 48} & \rightarrow & {\bf 7} \oplus {\bf 14} \oplus {\bf 27}.
\label{so7g2}
\eea
Combining the two results (\ref{so8so7}) and (\ref{so7g2}),
one obtains the following branching rule which is necessary to analyze
the section 2.

$\bullet$ $SO(8) \rightarrow G_2$  branching rule
\bea
{\bf 1} & \rightarrow & {\bf 1}, \nonu \\
{\bf 8} & \rightarrow & {\bf 1} \oplus {\bf  7}, \nonu \\
{\bf 28} & \rightarrow & 
{\bf 7} \oplus {\bf 7} \oplus 
{\bf 14}, \nonu \\
{\bf 56} & \rightarrow & {\bf 1} \oplus
{\bf 7} \oplus {\bf 7} \oplus 
{\bf 14}  \oplus
{\bf 27}, \nonu \\
{\bf 70}  & \rightarrow & 
{\bf 1} \oplus
{\bf 1} \oplus {\bf 7} \oplus 
{\bf 7}  \oplus
{\bf 27} \oplus {\bf 27}.
\label{so8g2}
\eea
It is better to present the following branching rule for the
correpondence between ${\cal N}=1$ and ${\cal N}=2$ critical points.

$\bullet$ $G_2 \rightarrow SU(3)$ branching rule
\bea
{\bf 1} & \rightarrow & {\bf 1}, \nonu \\
{\bf 7} & \rightarrow & {\bf 1} \oplus {\bf 3 } \oplus {\bf \bar{3}}, \nonu \\
{\bf 14} & \rightarrow & {\bf 3} \oplus {\bf \bar{3}} \oplus {\bf 8}, \nonu \\
{\bf 27} & \rightarrow & {\bf 1} \oplus {\bf 3} \oplus {\bf \bar{3}}
\oplus {\bf 6} \oplus {\bf \bar{6}} \oplus {\bf 8}. 
\label{g2su3}
\eea

\section{Explicit form for $\widetilde{A}_1$-tensor}

The $R^8$ coordinates $x_I$ in (\ref{ttilde}) are related to the ones
$X_I$ in (\ref{7metric}) as follows: 
\bea
X_1 & \equiv & \frac{1}{\sqrt{2}}(x_2-x_6), \qquad 
X_2 \equiv -\frac{1}{\sqrt{2}}(x_3-x_7), \nonu \\
X_3 & \equiv & \frac{1}{\sqrt{2}}(x_4-x_8), \qquad 
X_4 \equiv -\frac{1}{\sqrt{2}}(x_1-x_5), \nonu \\
X_5  & \equiv & \frac{1}{\sqrt{2}}(x_2+x_6), \qquad 
X_6 \equiv \frac{1}{\sqrt{2}}(x_3+x_7), \nonu \\
X_7  & \equiv & \frac{1}{\sqrt{2}}(x_4+x_8), \qquad 
X_8 \equiv \frac{1}{\sqrt{2}}(x_1+x_5) =\cos \theta
\nonu
\eea
and we list here for the expressions for 
$\widetilde{A}_1$-tensor in (\ref{a1tilde})
\bea
\widetilde{A}_1^{\,88} & = &
\frac{(1 + a + i \sqrt{-1 + a b})^3}{(2 + a + b)^{\frac{7}{2}}} \left[
8 + 16b - 8ab + 12b^2 - 12ab^2 + a^2b^2 + 
            4b^3 - 6ab^3 \right. \nonu \\
& + & b^4 + (a - b)(2 + a + b)(8 + a^2 + 
                  a(2 - 6b) + 
                  b(2 + b)) \cos^2 \theta 
+  \sqrt{-1 + ab} \nonu \\
&\times & \left.  (-4 i(1 + b)(2 + b(2 - a + b)) + 
                  4i(2 + a + b)(2 + a + a^2 + b - 
                        2ab + b^2) \cos^2 \theta) \right],
\nonu \\
\widetilde{A}_1^{\,mm} & = &
\frac{1}{\sqrt{2+a+b}} \left[ a(1+a)(b^2+(a-b)(a+b) \cos^2 \theta ) -
4(2+b)(-1+ab) X_m^2 \right. \nonu \\
& - & \left. i \sqrt{-1+ab}(a(b^2 +(a-b)(a+b)\cos^2
  \theta)+4(-2+(-1+a)b) X_m^2 ) \right],
\nonu \\
\widetilde{A}_1^{\,mn} & = &
\frac{4}{\sqrt{2+a+b}} \left[ -i(-2+(-1+a)b)\sqrt{-1+ab}-(2+b)(-1+ab) 
\right] X_m X_n, \quad m \neq n
\nonu \\
\widetilde{A}_1^{\,m8} & = &
\frac{4(1+ a +i \sqrt{-1+ab})^3}{(2+a+b)^{\frac{3}{2}}(a-b+2i\sqrt{-1+ab})}
\left[2-2ab+i(a-b)\sqrt{-1+ab}\right] X_m \cos \theta,
\label{a1}
\eea
where $m, n=1, 2, \cdots, 7$. Also one can obtain the full expressions
for the $\widetilde{A}_2$-tensor of the theory which we do not present
here. We have checked from these $\widetilde{A}_1$ and
$\widetilde{A}_2$-tensors that the scalar potential has a simple form
and it is given by $V=2a^3\left[a^2+b^2 +(a^2-b^2) \cos 2\theta \right]^2$.


\begin{thebibliography}{99}

\bibitem{Maldacena}
  J.~M.~Maldacena,
  Adv.\ Theor.\ Math.\ Phys.\  {\bf 2}, 231 (1998)
  [Int.\ J.\ Theor.\ Phys.\  {\bf 38}, 1113 (1999)]
  [arXiv:hep-th/9711200];
%
  E.~Witten,
  Adv.\ Theor.\ Math.\ Phys.\  {\bf 2}, 253 (1998)
  [arXiv:hep-th/9802150];
%
  S.~S.~Gubser, I.~R.~Klebanov and A.~M.~Polyakov,
  Phys.\ Lett.\  B {\bf 428}, 105 (1998)
  [arXiv:hep-th/9802109].

\bibitem{Warner83}
  N.~P.~Warner,
  Phys.\ Lett.\  B {\bf 128}, 169 (1983),
%
  Nucl.\ Phys.\  B {\bf 231}, 250 (1984).

\bibitem{AP}
  C.~Ahn and J.~Paeng,
  Nucl.\ Phys.\  B {\bf 595}, 119 (2001)
  [arXiv:hep-th/0008065].

\bibitem{AW}
  C.~Ahn and K.~Woo,
  Nucl.\ Phys.\  B {\bf 599}, 83 (2001)
  [arXiv:hep-th/0011121].

\bibitem{AI02}
  C.~Ahn and T.~Itoh,
  Nucl.\ Phys.\  B {\bf 627}, 45 (2002)
  [arXiv:hep-th/0112010],

\bibitem{AR99}
  C.~Ahn and S.~J.~Rey,
  Nucl.\ Phys.\  B {\bf 572}, 188 (2000)
  [arXiv:hep-th/9911199],

\bibitem{CPW}
  R.~Corrado, K.~Pilch and N.~P.~Warner,
  Nucl.\ Phys.\  B {\bf 629}, 74 (2002)
  [arXiv:hep-th/0107220].

\bibitem{JLP01}
  C.~V.~Johnson, K.~J.~Lovis and D.~C.~Page,
  JHEP {\bf 0110}, 014 (2001)
  [arXiv:hep-th/0107261].

\bibitem{BL0711}
  J.~Bagger and N.~Lambert,
  Phys.\ Rev.\  D {\bf 77}, 065008 (2008)
  [arXiv:0711.0955 [hep-th]];
%
  Phys.\ Rev.\  D {\bf 75}, 045020 (2007)
  [arXiv:hep-th/0611108];
%
  JHEP {\bf 0802}, 105 (2008)
  [arXiv:0712.3738 [hep-th]].

\bibitem{Gustavsson07}
  A.~Gustavsson,
  arXiv:0709.1260 [hep-th];
%
  JHEP {\bf 0804}, 083 (2008)
  [arXiv:0802.3456 [hep-th]].

\bibitem{GMR}
  J.~Gomis, G.~Milanesi and J.~G.~Russo,
  arXiv:0805.1012 [hep-th];
%
  S.~Benvenuti, D.~Rodriguez-Gomez, E.~Tonni and H.~Verlinde,
  arXiv:0805.1087 [hep-th];
%
  P.~M.~Ho, Y.~Imamura and Y.~Matsuo,
  arXiv:0805.1202 [hep-th].

\bibitem{ABJM}
  O.~Aharony, O.~Bergman, D.~L.~Jafferis and J.~Maldacena,
  arXiv:0806.1218 [hep-th].

\bibitem{BKKS}
  M.~Benna, I.~Klebanov, T.~Klose and M.~Smedback,
  arXiv:0806.1519 [hep-th].

\bibitem{Ahn08}
  C.~Ahn,
  arXiv:0806.1420 [hep-th].


\bibitem{GGY}
  D.~Gaiotto, S.~Giombi and X.~Yin,
  arXiv:0806.4589 [hep-th].

\bibitem{Agarwal}
  A.~Agarwal,
  arXiv:0806.4292 [hep-th].

\bibitem{Hanany:2008qc}
  A.~Hanany, N.~Mekareeya and A.~Zaffaroni,
  arXiv:0806.4212 [hep-th].

\bibitem{Armoni:2008kr}
  A.~Armoni and A.~Naqvi,
  arXiv:0806.4068 [hep-th].

\bibitem{Furuuchi:2008ki}
  K.~Furuuchi, S.~Y.~Shih and T.~Takimi,
  arXiv:0806.4044 [hep-th].

\bibitem{Larsson:2008ke}
  T.~a.~Larsson,
  arXiv:0806.4039 [hep-th].

\bibitem{Minahan:2008hf}
  J.~A.~Minahan and K.~Zarembo,
  arXiv:0806.3951 [hep-th].

\bibitem{IK}
  Y.~Imamura and K.~Kimura,
  arXiv:0806.3727 [hep-th].

\bibitem{FigueroaO'Farrill:2008jr}
  J.~Figueroa-O'Farrill,
  arXiv:0806.3534 [math.RT];
  arXiv:0805.4760 [math.RT].

\bibitem{Sochichiu:2008jm}
  C.~Sochichiu,
  arXiv:0806.3520 [hep-th].

\bibitem{Curtright:2008jj}
  T.~L.~Curtright, D.~B.~Fairlie and C.~K.~Zachos,
  arXiv:0806.3515 [hep-th].

\bibitem{Honma:2008jd}
  Y.~Honma, S.~Iso, Y.~Sumitomo and S.~Zhang,
  arXiv:0806.3498 [hep-th].

\bibitem{Caviezel:2008ik}
  C.~Caviezel, P.~Koerber, S.~Kors, D.~Lust, D.~Tsimpis and M.~Zagermann,
  arXiv:0806.3458 [hep-th].

\bibitem{NT}
  T.~Nishioka and T.~Takayanagi,
  arXiv:0806.3391 [hep-th].

\bibitem{Blau:2008bp}
  M.~Blau and M.~O'Loughlin,
  arXiv:0806.3255 [hep-th];
%
  arXiv:0806.3253 [hep-th].

\bibitem{Bhattacharya:2008bj}
  J.~Bhattacharya and S.~Minwalla,
  arXiv:0806.3251 [hep-th].

\bibitem{deMedeiros:2008bf}
  P.~de Medeiros, J.~Figueroa-O'Farrill and E.~Mendez-Escobar,
  arXiv:0806.3242 [hep-th].

\bibitem{Bergshoeff:2008ix}
  E.~A.~Bergshoeff, M.~de Roo, O.~Hohm and D.~Roest,
  arXiv:0806.2584 [hep-th].

\bibitem{Cecotti:2008qs}
  S.~Cecotti and A.~Sen,
  arXiv:0806.1990 [hep-th].

\bibitem{Ezhuthachan:2008ch}
  B.~Ezhuthachan, S.~Mukhi and C.~Papageorgakis,
  arXiv:0806.1639 [hep-th].

\bibitem{GRVV}
  J.~Gomis, D.~Rodriguez-Gomez, M.~Van Raamsdonk and H.~Verlinde,
  arXiv:0806.0738 [hep-th].

\bibitem{Pass}
  F.~Passerini,
  arXiv:0806.0363 [hep-th].

\bibitem{Park:2008qe}
  J.~H.~Park and C.~Sochichiu,
  arXiv:0806.0335 [hep-th].

\bibitem{Bandres:2008kj}
  M.~A.~Bandres, A.~E.~Lipstein and J.~H.~Schwarz,
  arXiv:0806.0054 [hep-th];
  JHEP {\bf 0805}, 025 (2008)
  [arXiv:0803.3242 [hep-th]].

\bibitem{Gustavsson:2008bf}
  A.~Gustavsson,
  arXiv:0805.4443 [hep-th].

\bibitem{FigueroaO'Farrill:2008zm}
  J.~Figueroa-O'Farrill, P.~de Medeiros and E.~Mendez-Escobar,
  arXiv:0805.4363 [hep-th].

\bibitem{Lin:2008qp}
  H.~Lin,
  arXiv:0805.4003 [hep-th].

\bibitem{Banerjee:2008pd}
  S.~Banerjee and A.~Sen,
  arXiv:0805.3930 [hep-th].

\bibitem{Hosomichi:2008jd}
  K.~Hosomichi, K.~M.~Lee, S.~Lee, S.~Lee and J.~Park,
  arXiv:0805.3662 [hep-th].

\bibitem{Li:2008ez}
  M.~Li and T.~Wang,
  arXiv:0805.3427 [hep-th].

\bibitem{Jeon:2008bx}
  I.~Jeon, J.~Kim, N.~Kim, S.~W.~Kim and J.~H.~Park,
  arXiv:0805.3236 [hep-th].

\bibitem{Song:2008bi}
  Y.~Song,
  arXiv:0805.3193 [hep-th].

\bibitem{Krishnan:2008zm}
  C.~Krishnan and C.~Maccaferri,
  arXiv:0805.3125 [hep-th].

\bibitem{Ho:2008ve}
  P.~M.~Ho, Y.~Imamura, Y.~Matsuo and S.~Shiba,
  arXiv:0805.2898 [hep-th].

\bibitem{Fuji:2008yj}
  H.~Fuji, S.~Terashima and M.~Yamazaki,
  arXiv:0805.1997 [hep-th].

\bibitem{Honma:2008un}
  Y.~Honma, S.~Iso, Y.~Sumitomo and S.~Zhang,
  arXiv:0805.1895 [hep-th].

\bibitem{Morozov:2008rc}
  A.~Morozov,
  arXiv:0805.1703 [hep-th];
  JHEP {\bf 0805}, 076 (2008)
  [arXiv:0804.0913 [hep-th]].
\bibitem{Ho:2008nn}
  P.~M.~Ho and Y.~Matsuo,
  arXiv:0804.3629 [hep-th].

\bibitem{Papadopoulos:2008gh}
  G.~Papadopoulos,
  arXiv:0804.3567 [hep-th];
  JHEP {\bf 0805}, 054 (2008)
  [arXiv:0804.2662 [hep-th]].

\bibitem{Gauntlett:2008uf}
  J.~P.~Gauntlett and J.~B.~Gutowski,
  arXiv:0804.3078 [hep-th].

\bibitem{Bergshoeff:2008cz}
  E.~A.~Bergshoeff, M.~de Roo and O.~Hohm,
  arXiv:0804.2201 [hep-th].

\bibitem{Ho:2008bn}
  P.~M.~Ho, R.~C.~Hou and Y.~Matsuo,
  arXiv:0804.2110 [hep-th].

\bibitem{Gran:2008vi}
  U.~Gran, B.~E.~W.~Nilsson and C.~Petersson,
  arXiv:0804.1784 [hep-th].

\bibitem{Distler:2008mk}
  J.~Distler, S.~Mukhi, C.~Papageorgakis and M.~Van Raamsdonk,
  JHEP {\bf 0805}, 038 (2008)
  [arXiv:0804.1256 [hep-th]].

\bibitem{Lambert:2008et}
  N.~Lambert and D.~Tong,
  arXiv:0804.1114 [hep-th].

\bibitem{VanRaamsdonk:2008ft}
  M.~Van Raamsdonk,
  JHEP {\bf 0805}, 105 (2008)
  [arXiv:0803.3803 [hep-th]].

\bibitem{Berman:2008be}
  D.~S.~Berman, L.~C.~Tadrowski and D.~C.~Thompson,
  arXiv:0803.3611 [hep-th].

\bibitem{MuPa}
  S.~Mukhi and C.~Papageorgakis,
  JHEP {\bf 0805}, 085 (2008)
  [arXiv:0803.3218 [hep-th]].


\bibitem{HLL}
  K.~Hosomichi, K.~M.~Lee and S.~Lee,
  arXiv:0804.2519 [hep-th];
%
  J.~Gomis, A.~J.~Salim and F.~Passerini,
  arXiv:0804.2186 [hep-th].

\bibitem{dN82}
  B.~de Wit and H.~Nicolai,
  Phys.\ Lett.\  B {\bf 108}, 285 (1982);
%
  B.~de Wit and H.~Nicolai,
  Nucl.\ Phys.\  B {\bf 208}, 323 (1982).

\bibitem{dNW}
  B.~de Wit, H.~Nicolai and N.~P.~Warner,
  Nucl.\ Phys.\  B {\bf 255}, 29 (1985).

\bibitem{AI02-1}
  C.~Ahn and T.~Itoh,
  Nucl.\ Phys.\  B {\bf 646}, 257 (2002)
  [arXiv:hep-th/0208137],

\bibitem{AW02}
  C.~Ahn and K.~Woo,
  Nucl.\ Phys.\  B {\bf 634}, 141 (2002)
  [arXiv:hep-th/0109010];

\bibitem{AW03}
  C.~Ahn and K.~Woo,
  JHEP {\bf 0311}, 014 (2003)
  [arXiv:hep-th/0209128],

\bibitem{MP}
  A.~Mauri and A.~C.~Petkou,
  arXiv:0806.2270 [hep-th].

\bibitem{Strassler98}
  M.~J.~Strassler,
  arXiv:hep-th/9810223.

\bibitem{Heidenreich}
  W.~Heidenreich,
  Phys.\ Lett.\  B {\bf 110} (1982) 461;
%
  L.~Gualtieri,
  arXiv:hep-th/0002116;
%
  B.~de Wit,
  arXiv:hep-th/0212245.

\bibitem{dN87}
  B.~de Wit and H.~Nicolai,
  Nucl.\ Phys.\  B {\bf 281}, 211 (1987).

\bibitem{GW}
  M.~Gunaydin and N.~P.~Warner,
  Nucl.\ Phys.\  B {\bf 248}, 685 (1984).

\bibitem{NW}
  H.~Nicolai and N.~P.~Warner,
  Nucl.\ Phys.\  B {\bf 259}, 412 (1985).

\bibitem{PW}
  C.~N.~Pope and N.~P.~Warner,
  Phys.\ Lett.\  B {\bf 150}, 352 (1985).

\bibitem{KW}
  A.~Khavaev and N.~P.~Warner,
  Phys.\ Lett.\  B {\bf 522}, 181 (2001)
  [arXiv:hep-th/0106032].

\bibitem{JLP00}
  C.~V.~Johnson, K.~J.~Lovis and D.~C.~Page,
  JHEP {\bf 0105}, 036 (2001)
  [arXiv:hep-th/0011166].

\bibitem{ps} 
R. Slansky, Phys. Rep. {\bf 79} (1981) 1;
J. Patera and D. Sankoff, Tables of Branching rules for
Representations of Simple Lie Algebras(L'Universit\'{e} de Montr\'{e}al, 
Montr\'{e}al, 1973);
W. MacKay and J. Petera, Tables of Dimensions, Indices and 
Branching Rules for representations of Simple Algebras(Dekker, New
York, 1981).

\end{thebibliography}
\end{document}